\documentclass[copyright]{eptcs}
\usepackage{amsmath}
\usepackage{amssymb}

\newtheorem{theorem}{Theorem}
\newcounter{cdef}
\newtheorem{definition}[cdef]{Definition}
\newcounter{cexample}
\newtheorem{lemma}[theorem]{Lemma}
\newtheorem{corollary}[theorem]{Corollary}

\title{Learning Residual Finite-State Automata\\Using Observation Tables}
\author{Anna Kasprzik \institute{FB IV, University of Trier \email{kasprzik@informatik.uni-trier.de}}}
\def\titlerunning{Learning Residual Finite-State Automata}
\def\authorrunning{Anna Kasprzik}
\begin{document}
\maketitle

\begin{abstract}
We define a two-step learner for RFSAs based on an observation table by using an algorithm for minimal DFAs to build a table for the reversal of the language in question and showing that we can derive the minimal RFSA from it after some simple modifications. We compare the algorithm to two other table-based ones of which one (by Bollig et al.\ \cite{july09}) infers a RFSA directly, and the other is another two-step learner proposed by the author. We focus on the criterion of query complexity.\\
{\bf Keywords:} Grammatical inference, residual languages, observation tables
\end{abstract}

\section{Introduction}
The area of grammatical inference tackles the problem of inferring a description of a formal language (a grammar, an automaton) from given examples or other kinds of information sources. Various settings have been formulated and quite a lot of learning algorithms have been developed for them. One of the best studied classes with respect to algorithmical learnability is the class of {\em regular languages}.

A significant part of these algorithms, of which Angluin's $L^*$ \cite{angluin} was one of the first, use the concept of an {\em observation table}. If a table fulfils certain conditions we can directly derive a deterministic finite-state automaton (DFA) from it, and if the information suffices this is the minimal DFA for the language in question.

In the worst case the minimal DFA has exponentially more states than a minimal NFA for a language $L$, and as for many applications a small number of states is desirable it seems worth to consider if we cannot obtain an NFA instead. Denis et al.\ \cite{RFSA} introduce special NFAs -- {\em residual finite-state automata (RFSAs)} -- where each state represents a residual language of $L$. Every regular language has a unique minimal RFSA. Denis et al.\ give several learning algorithms for RFSAs \cite{RFSANFA,RFSAlearn,RFSArevers}, which, however, all work by adding or deleting states in an automaton.

We define a two-step learner for RFSAs based on an observation table by using an algorithm for minimal DFAs to build a table with certain properties for the reversal of the language $L$ and showing that we can derive the minimal RFSA for $L$ from this table after some simple modifications. We compare the algorithm to two other table-based ones of which one is an incremental Angluin-style algorithm by Bollig et al.\ \cite{july09} which infers a RFSA directly, and the other is another two-step algorithm proposed below. The comparison mainly focuses on query complexity. We find that in theory the algorithm in \cite{july09} does not outperform the combination of known algorithms inferring the minimal DFA with the modifications we propose (although it is shown in \cite{july09} that their algorithm behaves better in practice).

%We study two-step algorithms hoping that modularity contributes to a clearer view of what is going on in algorithms that retrieve the syntactic equivalence classes of a regular language using an observation table. Also, by experimenting with the different forms of duality between (a) a language and its reversal, (b) prefixes and suffixes, (c) rows and columns, and (d) equivalence classes and residual languages, and their interconnections we would like to underline the universal usability of the concept of observation table as a means to execute and record such a retrieval process at the same time without having to deal with the idiosyncrasies of an underlying mechanism such as an automaton (for an extensive discussion concerning observation tables and other kinds of suitable representations also see \cite{balcazar}). % It is our ultimate goal to illuminate the different aspects of the area of Grammatical Inference from as many angles as possible and to clarify the notions used in it in order to eventually set it on a better, even more general theoretical foundation.

\section{Basic notions and definitions}\label{prels}
\begin{definition} An {\em observation table} is a triple $T=(S,E,obs)$ with $S,E\subseteq\Sigma^*$ finite, non-empty for some alphabet $\Sigma$ and $obs:S\times E\longrightarrow\lbrace0,1\rbrace$ a function with $obs(s,e) =1$ if $se\in L$, and $obs(s,e)=0$ if $se\notin L $. The {\em row} of $s\in S$ is $row(s):=\lbrace(e,obs(s,e))|e \in E \rbrace$, and the {\em column} of $e\in E$ is $col(e):=\lbrace(s,obs(s,e))|$ $s\in S\rbrace$.
$S$ is partitioned into two sets \textsc{red} and \textsc{blue} where $uv\in\textsc{red}\Rightarrow u\in \textsc{red}$ for $u,v\in\Sigma^*$ ({\em prefix-closedness}), and $\textsc{blue}:=\lbrace sa\in S\setminus \textsc{red}|s\in\textsc{red}$, $a\in\Sigma\rbrace$.\end{definition}

\begin{definition} Let $T=(S,E,obs)$ with $S=\textsc{red}\cup\textsc{blue}$. Two elements $r,s\in S$ are {\em obviously different} (denoted by $r <> s$) iff $\exists e\in E$ such that $obs(r,e)\neq obs(s,e)$.  $T$ is {\em closed} iff $\neg\exists s\in\textsc{blue}:\forall r\in\textsc{red}: r<>s$. $T$ is {\em consistent} iff $\forall s_1,s_2\in\textsc{red}$, $s_1a,s_2a\in S$, $a\in\Sigma:$ $row(s_1)=row(s_2)\Rightarrow row(s_1a)=row(s_2a)$. \end{definition}

\begin{definition}\label{aut} A {\em finite-state automaton} is a tuple $\mathcal{A}=(\Sigma,Q,Q_0,F,\delta)$ with finite input alphabet $\Sigma$, finite non-empty state set $Q$, set of start states $Q_0\subseteq Q$, set of final accepting states $F\subseteq Q$, and a transition function $\delta:Q\times\Sigma\longrightarrow2^Q$.

If $Q_0=\lbrace q_0\rbrace$ and $\delta$ maps at most one state to any pair in $Q\times\Sigma$ the automaton is {\em deterministic} (a {\em DFA}), otherwise {\em non-deterministic} (an {\em NFA}). If $\delta$ maps at least one state to every pair in $Q\times\Sigma$ the automaton is {\em total}, otherwise {\em partial}.

The transition function can always be extended to $\delta:Q\times\Sigma^*\longrightarrow2^Q$ defined by $\delta(q, \varepsilon)=\lbrace q\rbrace$ and $\delta(q,wa)=\delta(\delta(q,w),a)$ for $q\in Q$, $a\in\Sigma$, and $w\in \Sigma^*$.

Let $\delta(Q',w):=\bigcup\lbrace\delta(q,w)|q\in Q'\rbrace$ for $Q'\subseteq Q$ and $w\in\Sigma^*$. A state $q\in Q$ is {\em reachable} if there is $w\in\Sigma^*$ with $q\in\delta(Q_0,w)$. A state $q\in Q$ is {\em useful} if there are $w_1,w_2\in\Sigma^*$ with $q\in\delta(Q_0,w_1)$ and $\delta(q,w_2)\cap F\neq\emptyset$, otherwise {\em useless}.

The language accepted by $\mathcal{A}$ is $\mathcal{L}(\mathcal{A}):=\lbrace w\in\Sigma^*|\delta(Q_0,w)\cap F\neq  \emptyset\rbrace$.
\end{definition}

From $T=(S,E,obs)$ with $S=\textsc{red}\cup\textsc{blue}$ and $\varepsilon\in E$ derive an automaton $\mathcal{A}_T:=(\Sigma,Q_T,Q_{T0},F_T,\delta_T)$ defined by $Q_T=row(\textsc{red})$, $Q_{T0}=\lbrace row(\varepsilon)\rbrace$, $F_T=\lbrace row(s)|obs(s,\varepsilon)=1$, $s\in\textsc{red}\rbrace$, and $\delta_T(row(s),a)=\lbrace q\in Q_T|\neg(q<>row(sa))$, $s\in\textsc{red}$, $a\in\Sigma$, $sa\in S\rbrace$.
$\mathcal{A}_T$ is a DFA iff $T$ is consistent. The DFA for a regular language $L$ derived from a closed and consistent table has the minimal number of states (see \cite{angluin}, Th.\ 1). This DFA is the {\em canonical DFA $\mathcal{A}_L$ for $L$} and is unique.

The Myhill-Nerode equivalence relation $\equiv_L$ is defined by $r\equiv_L s\text{ iff }re\in L\Leftrightarrow se \in L\text{ for all }r,s,e\in\Sigma^*$. The {\em index} of $L$ is $I_L:=|\lbrace[s_0]_L|s_0\in \Sigma^*\rbrace|$ where $[s_0]_L$ is the equivalence class under $\equiv_L$ containing $s_0$.
\begin{theorem} {\em({\bf Myhill-Nerode theorem} -- see for example \cite{hopull})}\label{myhill}\\$I_L$ is finite $\Leftrightarrow$ $L$ can be recognized by a finite-state automaton $\Leftrightarrow$ $L$ is regular.\end{theorem}
\noindent $\mathcal{A}_L$ has exactly $I_L$ states, each of which represents an equivalence class under $\equiv_L$.

\begin{definition} The {\em reversal} $\overline{w}$ of $w\in\Sigma^*$ is defined inductively by $\overline{\varepsilon}:=\varepsilon$ and $\overline{aw}:=\overline{w}a$ for $a\in\Sigma$, $w\in \Sigma^*$. The reversal of $X\subseteq\Sigma^*$ is defined as $\overline{X}:=\lbrace\overline{w}| w\in X\rbrace$. The reversal of an automaton $\mathcal{A}=(\Sigma,Q,Q_0,F,\delta)$ is defined as $\overline{\mathcal{A}}:=(\Sigma,Q,F,Q_0, \overline{\delta})$ with $\overline{\delta}(q',w)=\lbrace q\in Q|q'\in\delta(q,w)\rbrace$ for $q'\in Q$, $w \in \Sigma^*$.\end{definition}

\begin{definition} The {\em residual language (RL)} of $L\subseteq\Sigma^*$ with regard to $w\in\Sigma^*$ is defined as $w^{-1}L:=\lbrace v\in\Sigma^*|wv\in L\rbrace$. A RL $w^{-1}L$ is called {\em prime} iff $\bigcup \lbrace v^{-1}L|v^{-1}L\subsetneq w^{-1}L\rbrace\subsetneq w^{-1}L$, otherwise {\em composed}. \end{definition} By Theorem \ref{myhill} the set of distinct RLs of a language $L$ is finite iff $L$ is regular. There is a bijection between the RLs of $L$ and the states of the minimal DFA $\mathcal{A}_L=(\Sigma,Q_L,\lbrace q_L\rbrace,F_L,\delta_L)$ defined by $\lbrace w^{-1}L\mapsto q'|w\in\Sigma^*,$ $\delta_L(q_L,w)=\lbrace q'\rbrace \rbrace$.

Let $L_q:=\lbrace w|\delta(q,w)\cap F\neq\emptyset\rbrace$ for a regular language $L\subseteq\Sigma^*$, some automaton $\mathcal{A}=(\Sigma,Q,Q_0,F,\delta)$ recognizing $L$, and $q\in Q$.
\begin{definition} A {\em residual finite-state automaton (RFSA)} is an NFA $\mathcal{A}=(\Sigma,Q,$ $Q_0,F, \delta)$ such that $L_q$ is a RL of $\mathcal{L}(\mathcal{A})$ for all states $q\in Q$.\end{definition}

\begin{definition} The {\em canonical RFSA $\mathcal{R}_L=(\Sigma,Q_R,$ $Q_{R0},F_R,\delta_R)$} for $L\subseteq \Sigma^*$ is defined by $Q_R=\lbrace w^{-1}L|w^{-1}L$ $\text{is prime}\rbrace$, $Q_{R0}=\lbrace w^{-1}L\in Q_R| w^{-1}L\subseteq L\rbrace$, $F_R=\lbrace w^{-1}L|\varepsilon\in w^{-1}L\rbrace$, and $\delta_R(w^{-1}L,a)=\lbrace v^{-1}L\in Q_R|$ $v^{-1}L\subseteq(wa)^{-1}L\rbrace$.\end{definition}
$\mathcal{R}_L$ is minimal with respect to the number of states (see \cite{RFSA}, Theorem 1).

\section{Inferring a RFSA using an observation table}\label{main}
\subsection{A ``parasitic'' two-step algorithm}\label{para}
The learner we define infers the canonical RFSA for $L$ from a suitable combination of information sources. A source can be an oracle for membership queries (MQs; `Is this string contained in the language?') or equivalence queries (EQs; `Is $A$ a correct automaton for $L$?' -- yielding some $c\in(L\setminus\mathcal{L}(A))\cup (\mathcal{L}(A)\setminus L)$ in case of a negative answer) or a positive or negative sample of $L$ fulfilling certain properties, and other kinds of sources can be considered as well. Suitable known combinations are: An oracle for MQs and EQs (a {\em minimally adequate teacher}, or MAT), an oracle for MQs with positive data, or positive and negative data.

In a first step we use an existing algorithm to build a table $T'=(\textsc{red}'\cup\textsc{blue}',E',obs')$ representing the canonical DFA for the reversal $\overline{L}$ of $L$. For eligible algorithms for various settings see \cite{angluin} ($L^*$, MAT learning), \cite{altex} (learning from MQs and positive data), or \cite{genmodel} (this meta-algorithm covers MAT learning, MQs and positive data, and positive and negative data, and can be adapted to other combinations). All these learners add elements to the set labeling the rows of a table (candidates for states in $\mathcal{A}_L$) until it is closed, and/or separating contexts (i.e., suffixes revealing that two states should be distinct) to the set labeling the columns until it is consistent -- additions of one kind potentially resulting in the necessity of the other and vice versa -- and, once the table is closed and consistent, deriving a DFA from it that is either $\mathcal{A}_L$ or can be rejected by a counterexample from the information sources, which is evaluated to restart the cycle. Obviously, since the sources only provide information about $L$ and not $\overline{L}$, we must minimally interfere by adapting data and queries accordingly: Strings and automata have to be reversed before submitting them to an oracle, samples and counterexamples before using them to construct $T'$.

In the second step we submit $T'$ to the following modifications:
\begin{itemize}
\item[$(1)$] Only keep one representative for every distinct row occurring in the table in $\textsc{red}'$, and only keep one representative for every distinct column in $E'$.
\item[$(2)$] Eliminate all representatives of rows and columns containing only $0$s.\\Let the resulting table be $T''=(\textsc{red}''\cup\textsc{blue}'',E'',obs'')$.
\item[$(3)$] Eliminate all representatives of {\em coverable} columns, i.e., all $e\in E''$ with \begin{itemize}\item[ ]$\exists e_1,\ldots,e_n\in E'':\forall s\in\textsc{red}'':$\\$[obs''(s,e)=0\Rightarrow \forall i\in \lbrace1,\ldots,n\rbrace: obs''(s,e_i)=0]$ $\wedge$\\$[obs''(s,e)=1\Rightarrow\exists i\in\lbrace 1,\ldots,n\rbrace:obs''(s,e_i)=1]$.\end{itemize}
For example, the column labeled by $e$ in Figure \ref{coverex} would be eliminated because its $1$s are all ``covered'' by the columns labeled by $e_1$, $e_2$, and $e_3$.
\begin{figure}{\centering $\begin{array}{c|ccccccccccc}&&e&&\text{ }e_1&&e_2&&e_3&&e_4\\\hline s_1&&1&&0&&1&&1&&0\\s_2&&1&&1&&0&&1&&1\\ s_3&&1&&0&&1&&0&&0\\s_4&&0&\text{ }&0&&0&&0&&1\\\end{array}$\\}
\caption{An example for a coverable column (labeled by $e$)}\label{coverex}\end{figure}\end{itemize} Note that the first two modifications mainly serve to trim down the table to make the third modification less costly. In fact, most algorithms mentioned above can easily be remodeled such that they build tables in which there are no rows or columns consisting of $0$s and in which the elements labeling the rows in the \textsc{red} part are pairwise obviously different already such that no row is represented twice.

The table thus modified shall be denoted by $T=(\textsc{red}\cup\textsc{blue},E,obs)$ and the derived automaton by $\mathcal{A}_T=(\Sigma,Q_T,Q_{T0},F_T,\delta_T)$ with $F_T=F_{T'}$ (this has to be stated in case $\varepsilon$ has been eliminated). As we have kept a representative for every distinct row and as all pairs of $\textsc{red}'$ elements that are distinguished by the contexts eliminated by $(3)$ must be distinguished by at least one of the contexts covering those as well $\mathcal{A}_T$ still represents $\mathcal{A}_{\overline{L}}$ (but without a failure state).

We use $T$ to define $\mathcal{R}:=(\Sigma,Q_R,$ $Q_{R0},F_R,\delta_R)$ with $Q_R=\lbrace q\subseteq\textsc{red}| \exists e\in E:s\in q\Leftrightarrow obs(s,e)=1\rbrace$, $Q_{R0}=\lbrace q\in Q_R|\forall s\in q:obs'(s, \varepsilon)=1\rbrace$ ($obs'$ in case $\varepsilon$ has been eliminated), $F_R=\lbrace q\in Q_R|\varepsilon\in q \rbrace$, and $\delta_R(q_1,a)=\lbrace q_2|q_2\subseteq\overline{\delta_T}(q_1,a)\rbrace$ for $q_1,q_2\in Q_R$ and $a\in\Sigma$, and $\overline{\delta_T}$ is the transition function of the reversal of $\mathcal{A}_{T}$.\\Observe that every state in $Q_R$ corresponds to a column in $T$. As every element of \textsc{red} represents an equivalence class of $\overline{L}$ under the Myhill-Nerode relation every state in $Q_R$ also corresponds to a unique set of equivalence classes, and the associated column represents the characteristic function of that set.

We show that $\mathcal{R}$ is the canonical RFSA for $L$. The proof uses Theorem \ref{theorem3}.
\begin{definition} Let $A=(\Sigma,Q,Q_0, F,\delta)$ be an NFA, and define $Q^\diamond:=\lbrace p\subseteq Q|\exists w\in\Sigma^*:\delta(Q_0,w)=p\rbrace$.\linebreak A state $q\in Q^\diamond$ is said to be {\em coverable} iff there exist $q_1,\ldots,q_n\in Q^\diamond\setminus\lbrace q\rbrace$ for $n\geq1$ such that $q=\bigcup^{n}_{i=1}q_i$.\end{definition}
\begin{theorem}{\em(Cited from \cite{RFSA}).}\label{theorem3} Let $L$ be regular and let $B=(\Sigma,Q_B,Q_{B0},F_B,$ $\delta_B)$ be an NFA such that $\overline{B}$ is a RFSA recognizing $\overline{L}$ whose states are all reachable. Then $C(B)=(\Sigma,Q_C,Q_{C0},F_C,\delta_C)$ with $Q_C=\lbrace p\in Q_B^\diamond|p\text{ is not coverable}\rbrace$, $Q_{C0}=\lbrace p\in Q_C|p\subseteq Q_{B0}\rbrace$, $F_C=\lbrace p\in Q_C|p\cap F_B\neq\emptyset\rbrace$, and $\delta_C(p,a)=\lbrace p'\in Q_C|p'\subseteq\delta_B(p,a)\rbrace$ for $p\in Q_C$ and $a\in\Sigma$ is the canonical RFSA recognizing $L$.\end{theorem}
As a further important result it has also been shown in \cite{RFSA}, Section 5, that in a RFSA for some regular language $L$ whose states are all reachable the non-coverable states correspond exactly to the prime RLs of $L$ and that consequently $Q_C$ can be identified with the set of states of the canonical RFSA for $L$.
\begin{lemma}{\em(See \cite{RFSA}, Prop.\ 1).}\label{prime} Let $A=(\Sigma,Q,Q_0,F,\delta)$ be a RFSA. For every prime RL $w^{-1}\mathcal{L}(A)$ there exists a state $q\in\delta(Q_0,w)$ such that $L_q=w^{-1}\mathcal{L}(A)$.\end{lemma}
\begin{theorem}\label{result} $\mathcal{R}$ is the canonical RFSA for $L$.\end{theorem}
{\bf Proof.} $\mathcal{A}_T$ meets the conditions for $\overline{B}$ in Theorem \ref{theorem3} as (a) all states of $\mathcal{A}_T$ are reachable because $\mathcal{A}_T$ contains no useless states, (b) $\mathcal{A}_T$ is a RFSA: Every DFA without useless states is a RFSA (see \cite{RFSA}), and (c) $\mathcal{L}(\mathcal{A}_T)=\overline{L}$.
As $\mathcal{A}_T$ contains no useless states $\mathcal{A}_T$ and $\overline{\mathcal{A}_T}$ have the same number of states and transitions, so we can set $B=\overline{\mathcal{A}_T}=(\Sigma,Q_T,F_T,Q_{T0}, \overline{\delta_T})$.
Assuming for now that there is indeed a bijection between $Q_R$ and $Q_C$ it is rather trivial to see that
\begin{itemize} \item there is a bijection between $Q_{R0}=\lbrace q\in Q_R|\forall x\in q:obs'(x,\varepsilon)=1 \rbrace$ and $Q_{C0}=\lbrace p\in Q_C|p\subseteq F_T\rbrace$ due to $F_T=\lbrace x\in\textsc{red}| obs'(x, \varepsilon)=1\rbrace$,
\item there is a bijection between $F_R=\lbrace q\in Q_R|\varepsilon\in q\rbrace$ and $F_C=\lbrace p\in Q_C|p\cap Q_{T0}\neq\emptyset\rbrace$ due to the fact that $Q_{T0}=\lbrace\varepsilon\rbrace$, and that
\item for every $q\in Q_R$, $p\in Q_C$, and $a\in\Sigma$ such that $q$ is the image of $p$ under the bijection between $Q_R$ and $Q_C$, $\delta_R(q,a)=\lbrace q_2\in Q_R|q_2\subseteq\overline{\delta_T}(q,a)\rbrace$ is the image of $\delta_C(p,a)=\lbrace p'\in Q_C|p'\subseteq\overline{\delta_T}(p,a)\rbrace$.\end{itemize}

It remains to show that there is a bijection between $Q_R$ and the set of prime RLs of $L$, i.e., $Q_C$. From the definition of $Q_R$ it is clear that $\mathcal{R}$ is a RFSA: As noted above, every state in $Q_R$ corresponds to a column in $T$, labeled by a context $e\in E$, and also to the set of equivalence classes $[s]_{\overline{L}}$ such that $se\in\overline{L}$ for $s\in\textsc{red}$. As a consequence the reversal of the union of this set of equivalence classes equals the RL $\overline{e}^{-1}L$, and hence every state in $Q_R$ corresponds to exactly one RL of $L$. According to Lemma \ref{prime}, there is a state in $Q_R$ for each prime RL of $L$, so every prime RL of $L$ is represented by exactly one column in $T$.\\By $(3)$ we have eliminated the columns that are covered by other columns in the table. If a column is not coverable in the table the corresponding state in $Q_R$ is not coverable either: Consider a column in the table which can be covered by a set of columns of which at least some do not occur in the table. Due to Lemma \ref{prime}, these columns can only correspond to composed RLs of $L$. If we were to add representatives of these columns to the table they would have to be eliminated again directly because of the restrictions imposed by $(3)$. This means that if a column is coverable at all it can always be covered completely by restricting oneself to columns that correspond to prime RLs of $L$ as well, and these are all represented in the table. Therefore $Q_R$ cannot contain any coverable states.\\Thus the correspondence between $Q_R$ and the set of prime RLs of $L$ is one-to-one, and we have shown that $\mathcal{R}$ is isomorphic to the canonical RFSA for $L$.\hfill$\blacksquare$

\begin{corollary} Let $L$ be a regular language. The number of prime RLs of $L$ is the minimal number of contexts needed to distinguish between the states of $\mathcal{A}_L$.\end{corollary}

Also note that we can skip the modification $(3)$ in the second part of our algorithm if we restrict the target to bideterministic regular languages (see \cite{angrevers}).

\subsection{Comparison to other algorithms: Query complexity}\label{comp}
An advantage of the algorithm described above is the trivial fact that it benefits from any past, present, and future research on algorithms that infer minimal DFAs via observation tables, and at least until now there is a huge gap between the amount of research that has been done on algorithms inferring DFAs and the amount of research on algorithms inferring NFAs -- or RFSAs, for that matter.

A point of interest in connection with the concepts presented is the study of further kinds of information sources that could be used as input and in particular suitable combinations thereof (see for example \cite{genmodel} for a tentative discussion).

Another point of interest is complexity. As the second part of the algorithm consists of cheap comparisons of $0$s and $1$s only of which $(3)$ is the most complex the determining factor is the complexity of the chosen underlying algorithm. One of the standard criteria for evaluating an algorithm is its time complexity, but depending on the different learning settings there are other measures that can be taken into consideration as well, one of which we will briefly address.

For algorithms that learn via queries a good criterion is the number of queries needed, obviously. The prototypical query learning algorithm, Angluin's \cite{angluin} algorithm $L^*$, which can be seen in a slightly adapted version $L^*_{col}$ in Figure \ref{lcol}, needs $O(I_L)$ equivalence queries and $O(|\Sigma|\cdot|c_0| \cdot I_L^2)$ membership queries, where $I_L$ is the index of $L\subseteq\Sigma^*$ and $|c_0|$ the length of the longest given counterexample. By modifications the number of MQs can be improved to $O(|\Sigma|I_L^2+I_Llog|c_0|)$ which according to \cite{balcazar} is optimal up to constant factors. On the other hand, it has been shown in \cite{tradeoff} that it is possible to decrease the number of EQs to sublinearity at the price of increasing the number of MQs exponentially.

\begin{figure}[ht!]
\begin{flushleft}{\tt {\bf initialize} $T:=(S,E,obs)$ with $S=\textsc{red}\cup\textsc{blue}$ and $\textsc{blue}= \textsc{red}\cdot\Sigma$\\
\hspace*{0.5cm} {\bf by} $\textsc{red}:=\lbrace\varepsilon\rbrace$ and $E:=\lbrace\varepsilon\rbrace$\\
{\bf repeat until} EQ $=$ yes\\
\hspace*{0.5cm} {\bf while} $T$ is not closed and not consistent\\
\hspace*{1cm} {\bf if} $T$ is not closed\\
\hspace*{1.5cm} find $s\in\textsc{blue}$ such that $row(s)\notin row(\textsc{red})$\\
\hspace*{1.5cm} $\textsc{red}:=\textsc{red}\cup\lbrace s\rbrace$ {\em (and update the table via MQs)}\\
\hspace*{1cm} {\bf if} $T$ is not consistent\\
\hspace*{1.5cm} find $s_1,s_2\in\textsc{red}$, $a\in\Sigma$, $e\in E$ such that $s_1a,s_2a\in S$\\
\hspace*{2cm} and $\neg(s_1<>s_2)$ and $obs(s_1ae)\neq obs(s_2ae)$\\
\hspace*{1.5cm} $E:=E\cup\lbrace ae\rbrace$ {\em (and update the table via MQs)}\\
\hspace*{0.5cm} {\em perform equivalence test}\\
\hspace*{0.5cm} {\bf if} EQ $=$ 0 get counterexample $c\in(L\setminus\mathcal{L}(\mathcal{A}_T))\cup (\mathcal{L}(\mathcal{A}_T)\setminus L)$\\
\hspace*{1cm} $E:=E\cup Suff(c)$ {\em (and update the table via MQs)}\\
{\bf return} $\mathcal{A}_T$} \end{flushleft}
\caption{$L^*_{col}$} \label{lcol}
\end{figure}

Recently, Bollig et al.\ \cite{july09} have presented a MAT algorithm for RFSAs using an observation table that keeps very close to the deterministic variant $L^*_{col}$ mentioned above. They introduce the notions of {\em RFSA-closedness} and {\em -consistency}.
\begin{definition}\label{defcov} Let $T=(S,E,obs)$ be an observation table. A row labeled by $s\in S$ is coverable iff $\exists s_1,\ldots,s_n\in S$ (is coverable by the rows of $s_1,\ldots,s_n$ iff) \begin{itemize} \item[ ] $\forall e\in E:[obs(s,e)=0\Rightarrow\forall i\in\lbrace1,\ldots,n\rbrace: obs(s_i,e)=0]$ $\wedge$\\ \hspace*{1.37cm} $[obs(s,e)=1\Rightarrow\exists i\in\lbrace1,\ldots,n\rbrace:obs(s_i,e)=1]$.\end{itemize} Let $ncov(S)\subseteq row(S)$ be the set of non-coverable rows labeled by elements in $S$.\end{definition}
\begin{definition}\label{defincl} Let $T=(S,E,obs)$ be an observation table. We say that a row $r\in row(S)$ {\em includes} another row $r'\in row(S)$, denoted by $r'\sqsubseteq r$, iff $obs(s',e)=1$ $\Rightarrow obs(s,e)=1$ for all $e\in E$ and $s,s'\in S$ with $row(s)=r$ and $row(s')=r'$.\end{definition}
\begin{definition}\label{defrclos} A table $T=(\textsc{red}\cup\textsc{blue},E,obs)$ is {\em RFSA-closed} iff every row $r\in row(\textsc{blue})$ is coverable by some rows $r_1,\ldots,r_n\in ncov(\textsc{red})$.\end{definition}
\begin{definition}\label{defrcons} A table $T=(\textsc{red}\cup\textsc{blue},E,obs)$ is {\em RFSA-consistent} iff $row(s_1)\sqsubseteq row(s_2)$ implies\linebreak$row(s_1a)\sqsubseteq row(s_2a)$ for all $s_1,s_2\in S$ and all $a\in\Sigma$.\end{definition}
From a RFSA-closed and -consistent table $T=(\textsc{red}\cup\textsc{blue},E,obs)$ Bollig et al.\ derive an NFA $\mathcal{R}=(\Sigma,Q_R,$ $Q_{R0},F_R,\delta_R)$ defined by $Q_R=ncov(\textsc{red})$, $Q_{R0}=\lbrace r\in Q_R|r\sqsubseteq row(\varepsilon)\rbrace$, $F_R=\lbrace r\in Q_R|\forall s\in\textsc{red}:row(s)=r\Rightarrow obs(s,\varepsilon)=1\rbrace$, and $\delta_R(row(u),a)=\lbrace r\in Q_R|r\sqsubseteq row(sa)\rbrace$ with $row(s)\in Q_R$ and $a\in\Sigma$.
\begin{theorem}{\em(See \cite{july09}).}\label{proofRFSA} Let $T$ be a RFSA-closed and -consistent table and $\mathcal{R}_T$ the NFA derived from $T$. Then $\mathcal{R}_T$ is a canonical RFSA for the target language. \end{theorem} See \cite{july09} for the proof. The algorithm $NL^*$ by Bollig et al.\ is given in Figure \ref{nlcol}.

\begin{figure}[ht!]
\begin{flushleft}{\tt {\bf initialize} $T:=(S,E,obs)$ with $S=\textsc{red}\cup\textsc{blue}$ and $\textsc{blue}= \textsc{red}\cdot\Sigma$\\
\hspace*{0.5cm} {\bf by} $\textsc{red}:=\lbrace\varepsilon\rbrace$ and $E:=\lbrace\varepsilon\rbrace$\\
{\bf repeat until} EQ $=$ yes\\
\hspace*{0.5cm} {\bf while} $T$ is not RFSA-closed and not RFSA-consistent\\
\hspace*{1cm} {\bf if} $T$ is not RFSA-closed\\
\hspace*{1.5cm} find $s\in\textsc{blue}$ such that $row(s)\in ncov(S)\setminus ncov(\textsc{red})$\\
\hspace*{1.5cm} $\textsc{red}:=\textsc{red}\cup\lbrace s\rbrace$ {\em (and update the table via MQs)}\\
\hspace*{1cm} {\bf if} $T$ is not RFSA-consistent\\
\hspace*{1.5cm} find $s\in S$, $a\in\Sigma$, $e\in E$ such that $obs(sae)=0$ and\\
\hspace*{2cm} $obs(s'ae)=1$ for some $s'\in S$ with $row(s')\sqsubseteq row(s)$\\
\hspace*{1.5cm} $E:=E\cup\lbrace ae\rbrace$ {\em (and update the table via MQs)}\\
\hspace*{0.5cm} {\em perform equivalence test}\\
\hspace*{0.5cm} {\bf if} EQ $=$ 0 get counterexample $c\in(L\setminus\mathcal{L}(\mathcal{A}_T))\cup (\mathcal{L}(\mathcal{A}_T)\setminus L)$\\
\hspace*{1cm} $E:=E\cup Suff(c)$ {\em (and update the table via MQs)}\\
{\bf return} $\mathcal{A}_T$} \end{flushleft}
\caption{$NL^*$, the NFA (RFSA) version of $L^*_{col}$} \label{nlcol}
\end{figure}

The theoretical query complexity of $NL^*$ amounts to at most $O(I_L^2)$ EQs and $O(|\Sigma|\cdot |c_0|\cdot I_L^3)$ MQs. This exceeds the maximal number of queries needed by $L^*_{col}$ in both cases which is due to the fact that with $NL^*$ adding a context does not always lead to a direct increase of the number of states in the automaton derived from the table. Note that the authors of \cite{july09} show that their algorithm statistically outperforms $L^*_{col}$ in practice, which is partly due to the fact that the canonical RFSA is often much smaller than the canonical DFA (see \cite{RFSA}). Nevertheless it is noteworthy that apparently inferring an automaton with potentially exponentially less states than the minimal DFA seems to be at least as complex.

Inspired by \cite{july09} we propose another parasitic two-step algorithm that uses an existing algorithm with access to a membership oracle to establish a table $T'=(\textsc{red}'\cup\textsc{blue}',E',obs')$ representing $\mathcal{A}_L$ and modifies it as follows:
\begin{itemize} \item[$(2)'$] Eliminate all representatives of rows and columns containing only $0$s. Let $T''=(\textsc{red}''\cup\textsc{blue}'',$ $E'',obs'')$ be the resulting table.
\item[$(3)'$] For every $s\in\textsc{red}''$ and every final state $q_F$ of $\mathcal{A}_{T''}$ add an (arbitrary) string $e$ to $E''$ such that $\delta_{T''}(row(s),e)=\lbrace q_F\rbrace$. Fill up the table via MQs. \end{itemize}
Let $T=(\textsc{red}\cup\textsc{blue},E,obs)$ be the resulting table. Note that as $T'$ already contains the maximal number of possible distinct rows $T$ is still closed and therefore RFSA-closed. $T$ is RFSA-consistent as well: Recall that every element $s\in S$ represents a RL $s^{-1}L$ of $L$ (see Section \ref{prels}). If $T$ was not RFSA-consistent we could find elements $s_1,s_2\in S$, $e\in E$, and $a\in\Sigma$ with $row(s_1)\sqsubseteq row(s_2)$ but $obs(s_1a,e)=1$ $\wedge$ $obs(s_2a,e)=0$. However, $ae\in s_1^{-1}L$ and $row(s_1)\sqsubseteq row(s_2)$ imply that $ae\in s_2^{-1}L$, and hence $obs(s_2a,e)=0$ cannot be true.

From $T$ we derive an automaton $\mathcal{R}=(\Sigma,Q_R,$ $Q_{R0},F_R,\delta_R)$ as in \cite{july09} (see above). The NFA $\mathcal{R}$ is the canonical RFSA for $L$. This follows directly from Theorem \ref{proofRFSA} and the fact that $T$ contains a representative for every RL of $L$.

The algorithm outlined above needs $I_L\cdot|F_L|$ MQs in addition to the queries needed by the algorithm establishing the original table but it does not require any more EQs. As EQs are usually deemed very expensive this can be counted in favor. Also note that if we restrict the target to bideterministic languages the table does not have to be modified and no additional queries have to be asked.

\section{Conclusion}\label{conclusion}
Two-step algorithms have the advantage of modularity: Their components can be exchanged and improved individually and therefore more easily adapted to different settings and inputs whereas non-modular algorithms are generally stuck with their parameters. One may doubt the efficiency of our two-step algorithms by observing that the second step partly destroys the work of the first, but as long as algorithms inferring the minimal DFA are much less complex than the ones inferring the minimal RFSA the two-step version outperforms the direct one.

It seems easy to adapt $NL^*$ to other learning settings such as learning from positive data and a membership oracle or from positive and negative data in order to establish a more universal pattern for algorithms that infer a RFSA via an observation table similar to the generalization for DFAs attempted in \cite{genmodel}.


\begin{thebibliography}{}
\bibitem{angluin} Angluin, D.: Learning regular sets from queries and counterexamples. Information and Computation 75(2), 87--106 (1987)
\bibitem{angrevers} Angluin, D.: Inference of reversible languages. JACM, vol. 29(3), pp. 741--765 (1982)
%\bibitem{dreweshoegberg} Drewes, F., H{\"o}gberg, J.: Learning a regular tree language from a teacher. In: DLT 2003. LNCS, vol. 2710, pp. 279--291. Springer (2003)
\bibitem{hopull} Hopcroft, J. E. and Ullmann, J. D.: Introduction to Automata Theory, Languages, and Computation. Addison-Wesley Longman (1990)
\bibitem{RFSA} Denis, F., Lemay, A., and Terlutte, A.: Residual Finite State Automata. In: STACS 2001. LNCS, vol. 2010, pp. 147--155. Springer (2001)
\bibitem{RFSANFA} Denis, F., Lemay, A., and Terlutte, A.: Learning regular languages using non-deterministic finite automata. In: ICGI. LNCS, vol. 1891, pp. 39--50. Springer (2000)
\bibitem{RFSAlearn} Denis, F., Lemay, A., and Terlutte, A.: Learning regular languages using RFSA. In: ALT 2001. LNCS, vol. 2225, pp. 348--363. Springer (2001)
\bibitem{RFSArevers} Denis, F., Lemay, A., and Terlutte, A.: Some classes of regular languages identifiable in the limit from positive data. In: Grammatical Inference -- Algorithms and Applications. LNCS, vol. 2484, pp. 269--273. Springer (2003)
\bibitem{july09} Bollig, B., Habermehl, P., Kern, C., and Leucker, M.: Angluin-style learning of NFA. In: Online Proceedings of IJCAI 21 (2009). %Available from: {\tt http://ijcai.org/papers09/contents.php}
%\bibitem{RFSAtrees} Carme, J., Gilleron, R., Lemay, A., Terlutte, A., and Tommasi, M.: Residual finite tree automata. In: DLT 2003. LNCS, vol. 2710, pp. 171--182. Springer (2003)
%\bibitem{BiRFSA} Latteux, M., Roos, Y., and Terlutte, A.: Minimal NFA and biRFSA languages. RAIRO Theoretical Informatics and Applications, vol. 43, pp. 221--237 (2009)
%\bibitem{BiRFSAlearn} Latteux, M., Lemay, A., Roos, Y., and Terlutte, A.: Identification of biRFSA languages. TCL, vol. 356(1), pp. 212--223 (2006)
\bibitem{altex} Besombes, J. and Marion, J.-Y.: Learning Tree Languages from Positive Examples and Membership Queries. In: ALT. LNCS, vol. 3244, pp. 440--453. Springer (2003)
\bibitem{balcazar} Balcazar, J.L., Diaz, J., Gavalda, R., and Watanabe, O.: Algorithms for learning finite automata from queries -- a unified view. In: Advances in Algorithms, Languages, and Complexity, pp. 53--72 (1997)
\bibitem{tradeoff} Balcazar, J.L., Diaz, J., Gavalda, R., and Watanabe, O.: The query complexity of learning DFA. New Generation Computing, vol. 12(4), pp. 337--358. Springer (1994)
%\bibitem{omega} Maler, O. and Pnueli, A.: On the learnability of infinitary regular sets. In: Proceedings of the 4th Annual Workshop on Computational Learning Theory, pp. 128--136. Morgan Kaufmann (1991)
\bibitem{genmodel} Kasprzik, A.: Meta-Algorithm GENMODEL: Generalizing over three learning settings using observation tables. Technical report 09-2, University of Trier (2009)
%\bibitem{fsmnlp08} Kasprzik, A.: Making finite-state methods applicable to languages beyond context-freeness via multi-dimensional trees. In J. Piskorski, B. Watson, A. {Yli-Jyr\"a} (eds): Post-proceedings of FSMNLP 2008, pp. 98--109. IOS Press (2009)
\bibitem{ndlstern} Kasprzik, A.: A learning algorithm for multi-dimensional trees, or: Learning beyond context-freeness. In A. Clark, F. Coste, L. Miclet (eds): ICGI 2008. LNAI, vol. 5278, pp. 111--124. Springer (2008)
\end{thebibliography}
\end{document}